\documentclass[a4paper,twoside]{article}
\newcommand{\affil}[1]{$^{\rm #1}$}
\usepackage[authoryear]{natbib}
\bibpunct{(}{)}{;}{a}{}{,}
\usepackage{graphicx}
\date{{\small Accepted for Publication 7 Sept 2010}} 
\title{\large\bf\flushleft Nucleosynthesis in the Stellar Systems $\omega$ Cen and M22
}
\author{\parbox{\textwidth}{\flushleft
\vspace{-0.5cm}
{\it G. S. Da Costa \affil{A,C} and A. F. Marino \affil{B}}\\
\vspace{0.4cm}
{\small \affil{A}\,Research School of Astronomy \& Astrophysics, Australian National University,
Mt Stromlo Observatory, via Cotter Rd, Weston, ACT 2611, Australia}\\
{\small \affil{B}\,Dipartimento di Astronomia, Universit\`{a} di Padova, vicolo dell'Osservatorio 2, I-35122
Padova, Italy}\\
{\small \affil{C}\,Email: gdc@mso.anu.edu.au}}}
\begin{document}
\twocolumn[
\begin{changemargin}{.8cm}{.5cm}
\begin{minipage}{.9\textwidth}
\vspace{-1cm}
\maketitle
%
%
{\small{\bf Abstract:}\\
The stellar system $\omega$ Centauri is well known for the large range in abundance among its 
member stars.  Recent work has indicated that the globular cluster M22 (NGC~6656) also possesses
an internal abundance range, albeit substantially smaller than that in $\omega$ Cen.  Here we 
compare, as a function of [Fe/H], element-to-iron ratios in the two systems for a number 
of different elements using data from abundance analyses of red giant branch stars.  It appears that
the nucleosynthetic enrichment processes were very similar in these two systems despite the substantial
difference in total mass. }

\medskip{\bf Keywords:} globular clusters: general --- globular clusters: individual (M22, $\omega$
Cen) --- stars: abundances

\medskip
\medskip
\end{minipage}
\end{changemargin}
]
\small

\section{Introduction}
The stellar system $\omega$ Cen has been known for a long time to be different from most
other Galactic globular clusters because its member stars exhibit a wide range in the abundance
of the heavier elements, such as Iron and Calcium.  This contrasts with the situation in most other globular 
clusters where the size of any intrinsic spread in the abundance of such elements is smaller than the 
measurement uncertainties \citep[e.g.,][]{KI03,CB09}.   The wide range in heavy element 
abundance\footnote{Usually represented by [Fe/H], the logarithm of the Iron abundance relative 
to that of the Sun.} in $\omega$ Cen was first demonstrated by abundance measures 
from spectra of cluster RR~Lyrae variables \citep{FR75}.  This then led to the recognition that the 
large colour-width of the red giant branch in the colour-magnitude diagram (CMD) of the cluster,
discovered by \citet{Wo66} and \citet{Ge67} and confirmed by \citet{CS73}, was a natural consequence
of the heavy abundance spread.  There now exists a large body of spectroscopic 
data that characterises the range in [Fe/H] in $\omega$ Cen, and more particularly, reveals the 
complex variations in [element/Fe] ratios with [Fe/H].  The element ratios reveal the nucleosynthetic 
history of the stellar system, and contributions from both Type II and Type Ia supernovae
are recognisable, as is a significant role for asymptotic giant branch (AGB) stars in the chemical evolution.

The difference between $\omega$ Cen and other globular clusters has led to the suggestion
that $\omega$ Cen may have formed in a different way from the other clusters --- that it is the
nuclear remnant of a former dwarf galaxy that has been tidally disrupted by the Milky Way
\citep[e.g.,][]{KF93}.  \citet{BF03}, for example, have shown that this is dynamically plausible.
The different evolutionary environment may then have allowed additional chemical processes
that do not occur in `regular' globular clusters \citep[e.g.,][]{BN06, Ro07,Ro10}.

The stellar system $\omega$ Cen is, however, no longer the only cluster in which a definite range in heavy
element abundance is present among the member stars.  For example, following on initial photometric studies of
\citet{SL95}, \citet{BI08} have used spectra at the Ca II triplet of a large sample of member stars to
show that the globular cluster M54, which is located at the centre of the Sagittarius dwarf galaxy, 
possesses an internal abundance range characterised by $\sigma$([Fe/H])$_{int}$ $\approx$ 0.14 dex
\citep[see also the very recent work of][]{CB10c}.
Further, two other globular clusters have recently been shown to have internal ranges in heavy element
abundance.  The first of these is the Galactic bulge globular cluster Terzan~5, for which observations 
presented
in \citet{FD09}  reveal the presence of two distinct populations that differ by a factor of $\sim$3 in
[Fe/H], and which likely also differ significantly in age, with the more metal-rich population being 
younger \citep{FD09}.  The second globular cluster is M22 (NGC~6656), a stellar system long suspected of
possessing similar properties to $\omega$~Cen \citep[e.g.,][]{NF83}, despite the fact that with 
M$_V$ $\approx$ \mbox{--8.5}, it is considerably less luminous than either $\omega$ Cen or M54, which
have M$_V$ $\approx$ --10.3 and --10.0, respectively \citep{WH96}.  For M22, \citet{DH09}
used spectra at the Ca II triplet of 41 red giants to derive a [Fe/H] abundance distribution whose 
significant width could not be explained by differential reddening effects.  The inter-quartile range for the 
observed abundances is 0.24 dex, and the distribution suggests the presence of at least two 
components, whose mean metallicities differ by 0.26 dex \citep{DH09}.   
An intrinsic variation in star-to-star abundances in M22 was also revealed recently by
the high resolution spectroscopic study of 17 M22 red giants presented in \citet{AM09}.  
These authors found that, not only is there an intrinsic range in [Fe/H] of order 0.15 dex, but also 
two groups of stars are present in the cluster, with the Iron-richer stars having higher Calcium 
and $s$-process element abundances relative to the Iron-poorer stars \citep{AM09}.

While the number of clusters showing internal heavy element abundance variations is small,
this is not the case for the lighter elements.  After many decades of work, the light elements
C, N, O, Na, Al, and Mg are now known to vary significantly and systematically within all globular clusters 
for which adequate data are available \citep[e.g.,][]{RG04,CB10}.  The sense of the variations is such
that increases in the Nitrogen, Sodium and Aluminium abundances are coupled with decreases in the
Carbon, Oxygen and Magnesium abundances.  Together the effects are referred to as the 
O-Na anti-correlation.  The abundance patterns seen in the Na-rich, O-poor stars are consistent with being 
produced by the operation of the CNO-cycle, together with proton-capture reactions on Ne and Mg 
seeds during 
H-burning at high temperatures \citep{DD90, La93}.  However, the actual site of the nucleosynthesis
is not yet clearly established, with the most likely candidates being intermediate-mass AGB stars
\citep[e.g.,][]{DV07} or rapidly rotating massive stars \citep[e.g.,][]{De07}.  Nevertheless, given that the `anomalies' are found among
main sequence stars in at least some clusters, the origin of the abundance patterns must be related to a process or processes that
occurred during the formation of the clusters.  In this respect $\omega$ Cen and M22 are no
different from other globular clusters -- the O-Na anti-correlation is clearly present in both 
systems \citep{ND95a, AM09}. 

The stellar system $\omega$ Cen has one further property that distinguishes it from most other
globular clusters: the likely presence of a large range in Helium abundance.  This property is 
inferred from the presence of a double main-sequence in a high precision Hubble Space 
Telescope-based CMD for the cluster \citep[][see also \citet{BB10}]{Bd04}.  
In this CMD the bluer sequence is 
less numerous and of higher abundance than the redder sequence \citep{Po05}.  The observations 
are best interpreted as indicating that the stars in the bluer sequence are enhanced in Helium 
by $\Delta$Y $\approx$ 0.10 -- 0.15 relative to those in the redder sequence
\citep{JN04, Po05}.   Such large He abundance ranges have also been suggested to
occur in a small number of other luminous, massive globular clusters.  The prime example is
NGC~2808, whose CMD shows a triple main sequence \citep{Pi07}.  Given the lack of
heavy element abundance variation in the cluster \citep[e.g.,][]{CB09}, the main sequence structure
is best interpreted as indicating distinct He abundance groups, which are then likely also related
to the multi-modal structure of the horizontal branch in the cluster CMD \citep[e.g.,][]{FDA05}.
Other clusters for which very precise CMDs reveal the existence of main sequence colour widths
in excess of that expected from the photometric errors include 47~Tuc \citep{AP09} and NGC~6752
\citep{MP10}.
The origin of the postulated large Y variations in these clusters is poorly understood though the same 
candidates as those employed for explaining the O-Na anti-correlation are often invoked 
\citep[e.g.,][]{AR08}.
However, although some structure is present in the vicinity of the subgiant branch
in an $HST$-based CMD for M22 \citep{GP09}, there is no evidence for any substantial He abundance 
variations within this cluster.  In this respect then, M22 clearly differs from $\omega$ Cen.

In this paper we concentrate on a comparison of the element-to-iron abundance ratios, as a function
of [Fe/H], between the red giants in $\omega$ Cen and M22.  This allows an investigation of the 
extent to which similar nucleosynthetic activities occurred during their evolution, which in turn, can
constrain their origin.  The observational
data used for the comparison are outlined in the next section.  The comparison is carried out in
Section 3, and the results are discussed in Section 4.  Briefly, a considerable degree of similarity is
found between the two systems, suggesting they underwent analogous evolutionary processes. 

\section{Observational Data}

There are a number of abundance analyses based on high-resolution spectroscopy available for
$\omega$ Cen red giants.  These include \citet{ND95b}, \citet{VS00}, \citet{EP02} and \citet{JP09}, 
each of which 
differ in sample size, elements studied, and metallicity range covered.  We will use primarily the
results of \citet{ND95b}, who give abundances for a large number of elements derived from
observations of a sample of 40 red giants
chosen to cover (almost) the full range of [Fe/H] values exhibited by $\omega$ Cen stars.  Where
necessary we will also draw on the results of \citet{VS00} and \citet{JP09}: both these papers have
demonstrated that any systematic differences between their results and those of \citet{ND95b} are
small, and have their origin in different choices of $gf$-values, lines measured, and atmospheric
parameters\footnote{After this manuscript was submitted, a paper by \citet{JP10} appeared on the
arXiv preprint archive.  The paper provides O, Na, Al, Si, Ca, Sc, Ti, Fe, Ni, La and Eu abundances 
for a very large sample of 855 $\omega$~Cen red giants.  We have attempted to include results
from that paper where appropriate, but note that \citet{JP10} show that any systematic offset between
their abundances and those of \citet{ND95b} are also small.}.

In contrast to $\omega$ Cen, the only published high dispersion study of a large sample of
M22 red giants is that of \citet{AM09}.  These authors analysed high-resolution VLT/UVES observations 
of 17 M22 red giants as well as somewhat lower resolution VLT/GIRAFFE data for a further 14 red 
giants, one of which is in common with the UVES sample.  

\begin{figure}
\begin{center}
\includegraphics[scale=0.42,angle=0.]{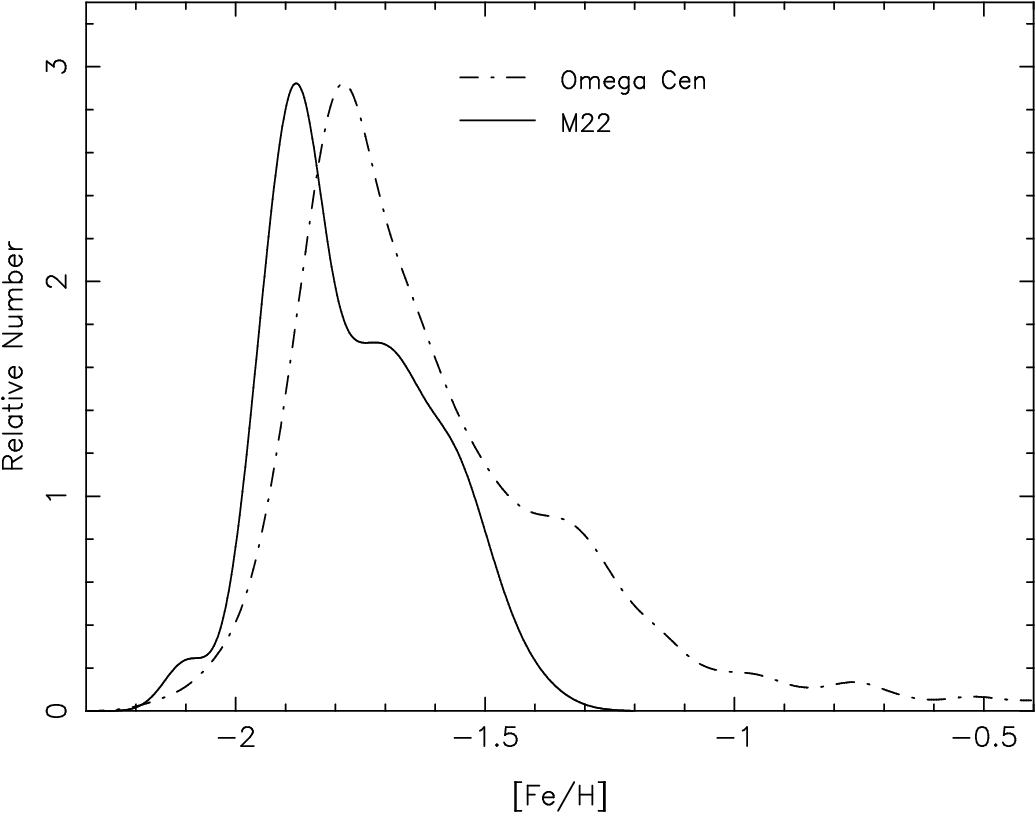}
\caption{Generalised histograms showing the abundance distributions of $\omega$ Cen
(dot-dash line) and M22 (solid line).  The two distributions have been normalised to the
same peak height.  See text for details.}
\label{ghisto_fig}
\end{center}
\end{figure}

\section{Comparing $\omega$ Cen and M22 Element-to-Iron Abundance Ratios}

To set the scene we show in Figure \ref{ghisto_fig} generalised histograms for the abundance 
distributions in M22 and $\omega$ Cen.  The M22 data are from \citet{DH09} while the [Fe/H]
distribution for $\omega$ Cen is derived from the [Ca/H] distribution presented in \citet{NFM96},
by assuming a constant [Ca/Fe] ratio of 0.4 dex \citep[][see also Figure \ref{sica_fig}]{ND95b}.
The two distributions, which incorporate similar abundance uncertainties \citep[see][]{DH09} have
been scaled to have the same peak height.  As noted by \citet{DH09}, the distributions show a
considerable degree of similarity.  Both rise rapidly on the metal-poor side to a well defined peak, with
that for M22 being $\sim$0.09 dex more metal-poor than that for $\omega$ Cen.  Both
distributions then decrease with increasing metallicity and there are clear hints for the presence of 
multiple populations.  The $\omega$ Cen distribution continues to substantially higher abundances
than that for M22, a result which is most likely not due solely to the smaller sample size for M22
\citep{DH09}.

We now turn to a comparison of element-to-iron abundance ratios between these two stellar
systems, starting first with the $\alpha$- and iron-peak elements, which constrain the contributions
from Type II and Type Ia supernovae.   This is followed by the elements O, Na and Al, which are
involved in the poorly understood globular-cluster-specific nucleosynthesis process, and finally
by the $r$- and $s$-process elements, the latter of which is closely tied to nucleosynthetic
contributions from AGB stars.

\subsection{The $\alpha$- and iron-peak elements}

\begin{figure}
\begin{center}
\includegraphics[scale=0.42,angle=0.]{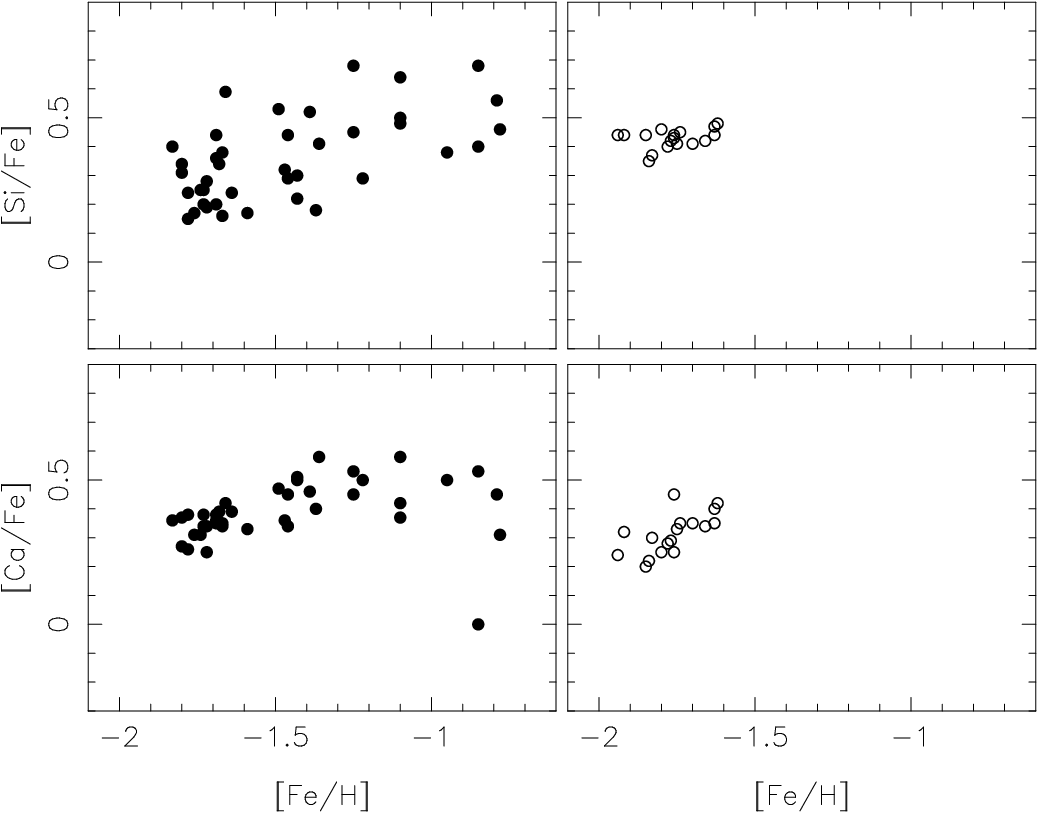}
\caption{Element-to-Iron abundance ratios for the $\alpha$-elements Silicon
(upper) and Calcium (lower panels) as a function of
[Fe/H] for $\omega$ Cen red giants (left panels) and for M22 red giants (right panels).  
Unless otherwise noted, in this and subsequent figures the $\omega$ Cen data are from
\citet{ND95b} and the M22 data are from \citet{AM09}. }
\label{sica_fig}
\end{center}
\end{figure}

In Figure \ref{sica_fig} we show element-to-iron ratios as a function of [Fe/H] for the 
$\alpha$-elements Si and Ca.  We note that the [Si/Fe] ratio for the $\omega$ Cen stars shows a sizeable
scatter with an indication of a possible trend of increasing [Si/Fe] with [Fe/H].  The [Si/Fe] data
for the relatively small number of red giants studied in \citet{VS00} show no such trend, but
the large sample of stars studied in \citet{JP10} reveals that the behaviour of [Si/Fe]
with [Fe/H] in $\omega$~Cen is quite complex.   The bulk of the population has [Si/Fe] = 0.29 but
the ratio rises to [Si/Fe] = 0.45 for the most metal-rich stars, consistent with the \citet{ND95b} data
shown in the Figure.
Similarly, the \citet{ND95b} [Ca/Fe] data plotted in the lower left panel of Figure \ref{sica_fig} show
some indication of a slight increase in [Ca/Fe] with [Fe/H] for the stars more metal-poor than 
[Fe/H] $\approx$ --1.3 dex.  The large sample study of \citet{JP10}  reveals similar behaviour, in that while
the bulk of the population has [Ca/Fe] = 0.26, there is a small ($\sim$0.1 dex) increase in the abundance 
ratio as [Fe/H] rises.  However, 
the stars with [Fe/H] $>$ --1 possess similar [Ca/Fe] values to that for 
the bulk of the population \citep{JP10}.
On the other hand,  for M22 there is clearly no trend in [Si/Fe] with [Fe/H], and the dispersion in the 
[Si/Fe] values is 
particularly small.  However, for [Ca/Fe], there may also be a small increase in [Ca/Fe] with [Fe/H] 
similar to that seen for $\omega$ Cen.  Further investigation is needed to confirm the
existence of this possible trend.

Similar plots for the other $\alpha$-elements for which abundances are available, Mg and Ti, show 
the same result  -- the red giants in both clusters have essentially constant and positive [Mg/Fe] 
and [Ti/Fe] values regardless of 
[Fe/H]\footnote{The extensive sample of \citet{JP10} shows that [Ti/Fe] behaves similarly to [Si/Fe]
and [Ca/Fe] in that there is a slight increase in [Ti/Fe] with increasing [Fe/H].}.  The one exception, as noted in \citet{ND95b}, is that the $\omega$ Cen sample shows a small 
number of Mg-depleted stars, with [Mg/Fe] values in the range $\sim$0.05 to --0.2 dex.  These stars, 
however, are all strongly enhanced in [Al/Fe] \citep{ND95b} presumably reflecting a significant 
contribution to the gas from which they formed from material processed through the Mg-Al cycle. 
No such Mg-depleted stars are found in the current M22 sample \citep{AM09}.

Consequently, in general it appears that both $\omega$~Cen and M22 show a similar degree
of enrichment from Type II SN, with the value of [$\alpha$/Fe] also being closely similar to 
that for Galactic halo field stars at comparable [Fe/H] values \citep[e.g.,][]{RG04}.
However, we note in passing that according to \citet{EP02}, at the 
highest [Fe/H] values (not shown in Figure \ref{sica_fig}), the red giants in $\omega$ Cen have lower 
[$\alpha$/Fe] values.  In particular, \citet{EP02} have analyzed three $\omega$ Cen
red giants with [Fe/H] $\approx$
--0.6, finding [$\alpha$/Fe] = 0.10 $\pm$ 0.04 ($\alpha$ = Ca, Si) compared to three red giants at
[Fe/H] $\approx$ --1.0 for which [$\alpha$/Fe] = 0.29 $\pm$ 0.01, a value comparable to those
seen in Figure \ref{sica_fig}.  The lower [$\alpha$/Fe] value at larger [Fe/H] is taken as an indication 
that Type Ia SN have contributed to the chemical evolution of $\omega$ Cen.  However, \citet{JP10}
have also analysed two of the three metal-rich stars studied in \citet{EP02}, and find higher [Si/Fe]
and [Ca/Fe] abundance ratios.  Whether this is solely the result of differences in adopted $gf$-values,
model atmospheres and lines measured \citep{JP10}, or is a more fundamental discrepancy, is
unclear.
The M22 sample lacks 
any low [$\alpha$/Fe] stars, though this is perhaps not surprising if one supposes that the smaller total 
abundance range in M22 means that the duration of the chemical evolution epoch was shorter in
that cluster than for $\omega$ Cen.

\begin{figure}
\begin{center}
\includegraphics[scale=0.42,angle=0.]{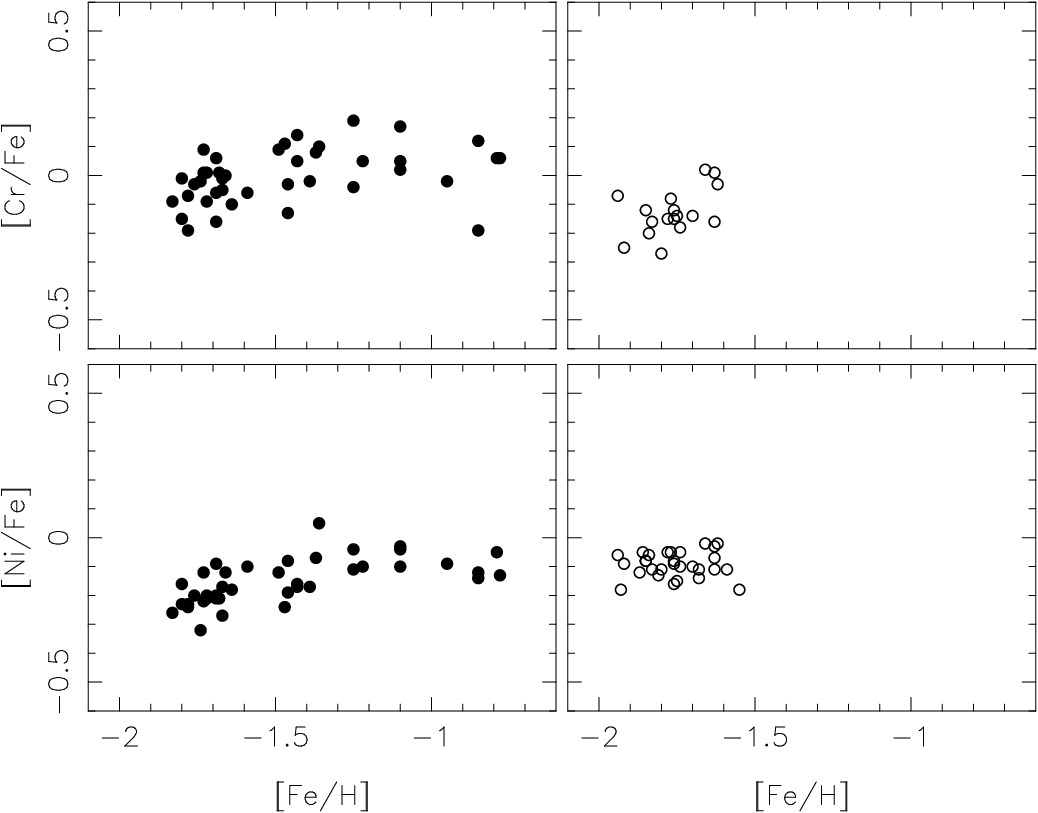}
\caption{Element-to-iron abundance ratios for the iron-peak elements Chromium
(upper) and Nickel (lower panels) as a function of
[Fe/H] for $\omega$ Cen red giants (left panels) and for M22 red giants (right panels).  
}
\label{crni_fig}
\end{center}
\end{figure}

In Figure \ref{crni_fig} we show [Cr/Fe] and [Ni/Fe] as a function of [Fe/H] for the $\omega$ Cen and 
M22 red giant samples.  Plots of the abundance ratios for Scandium and Vanadium, the other iron-peak 
elements for which data are available, show similar behaviour -- no dependence on [Fe/H] and 
ratio values near solar -- and the extensive data of \citet{JP10} for [Ni/Fe] and [Sc/Fe]  show the same
result.  Further, as for the $\alpha$-elements, the abundance ratios for the $\omega$ Cen and 
M22 red giants are similar to those for halo field stars at similar [Fe/H] values \citep[e.g.,][]{RG04}.
The similarity between the two stellar systems and the halo field is not surprising since at these
metallicities the production of iron-peak elements like Cr and Ni is expected to closely follow that of Fe.

\subsection{Oxygen, Sodium and Aluminium}

\begin{figure}
\begin{center}
\includegraphics[scale=0.42,angle=0.]{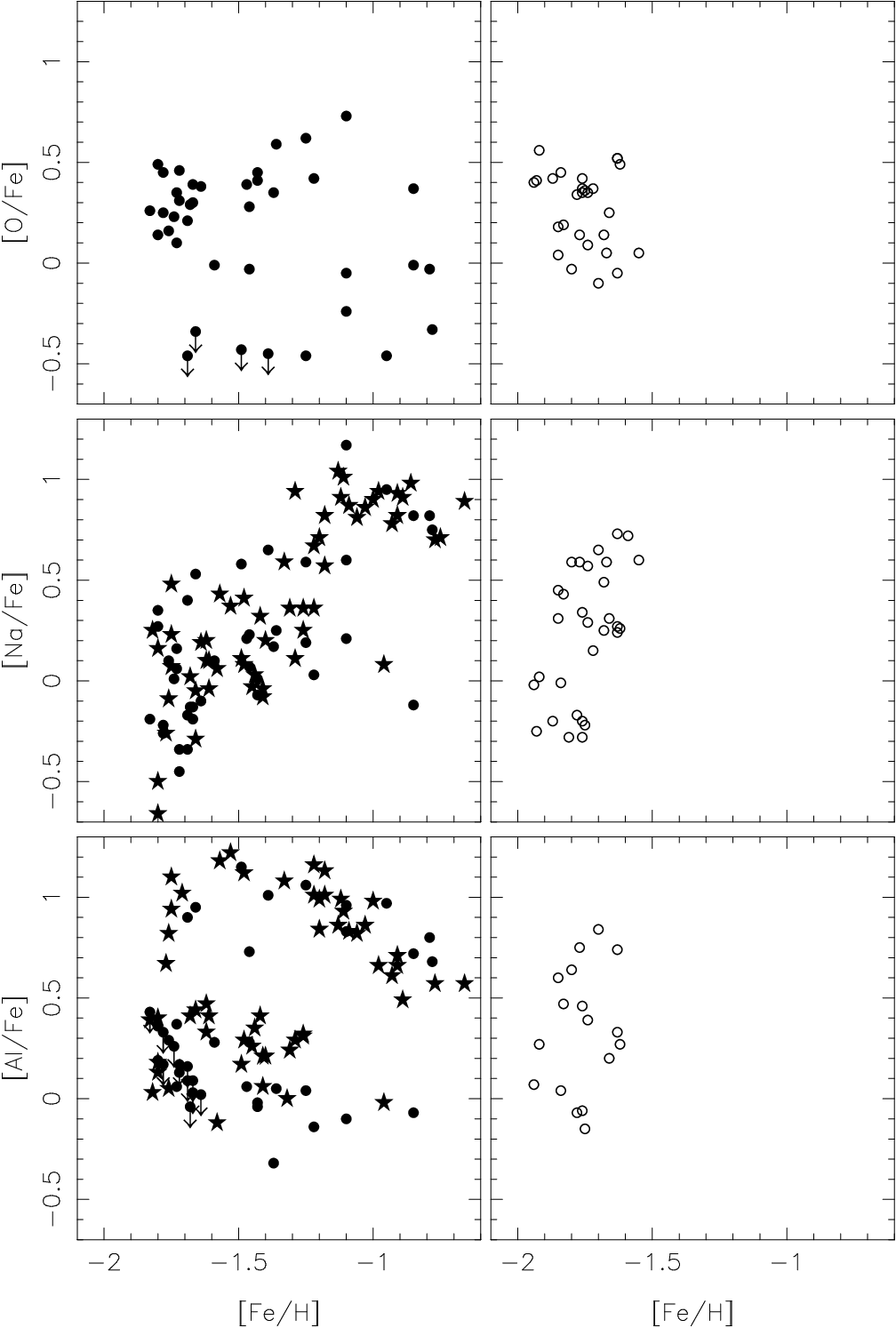}
\caption{Element-to-Iron abundance ratios for the  elements Oxygen
(upper), Sodium (middle) and Aluminium (lower panels) as a function of
[Fe/H] for $\omega$ Cen red giants (left panels) and for M22 red giants (right panels). For $\omega$
Cen, the filled circles are from \citet{ND95b} while the filled stars are from \citet{JP09}. 
$\omega$~Cen points with downward arrows represent upper limits.}
\label{onaal_fig}
\end{center}
\end{figure}

In Figure \ref{onaal_fig} we show the Oxygen, Sodium and Aluminium abundance ratios with respect 
to Iron for  $\omega$~Cen (left panels) and M22 (right panels).  The $\omega$ Cen data for [O/Fe]
come from \citet{ND95b} while that for [Na/Fe] and [Al/Fe] come from both \citet{ND95b} and
\citet{JP09}.  There are 7 stars in common.    For these 7 stars, the mean difference in [Na/Fe], in the 
sense of \citet{JP09} $minus$ \citet{ND95b} is --0.11 dex with a sigma of 0.20 dex.  For [Al/Fe],
\citet{ND95a} give only upper limits for two of the common stars, but for the remaining 5 the 
mean difference in [Al/Fe] is +0.08 dex with a sigma of 0.14 dex.  Thus any systematic difference
between the two sets of  abundance determinations is minor, and can be neglected given the 
substantial range in [Na/Fe] and [Al/Fe] observed.  The more extensive $\omega$~Cen data set of
\citet{JP10} shows essentially the same structure as seen in the left panels of Figure \ref{onaal_fig}.

Turning first to
Oxygen, we see that the upper limit for the [O/Fe] values, namely [O/Fe] $\approx$ 0.4--0.5, is similar in both clusters and shows no obvious trend with [Fe/H].  This is not surprising since the [O/Fe] ratio
for the ``O-rich'' stars is expected to be similar to the element-to-iron abundance ratio for other 
$\alpha$-elements (e.g., Figure \ref{sica_fig}).
Nevertheless, as is common among globular cluster stars, both $\omega$ Cen and M22 also exhibit 
populations of red giants with depleted [O/Fe] ratios, which correlate with enhanced Na and Al 
abundances \citep[e.g.,][]{ND95a,AM09}.  Specifically for the  [O/Fe] ratios, the only noteworthy 
difference between the two stellar systems, within the range of [Fe/H] overlap, is
the presence in $\omega$ Cen of a small number of stars showing very low [O/Fe] values; such stars 
are not seen in the M22 sample of \citet{AM09}.  Further, 
it is also evident from the upper left panel of Figure \ref{onaal_fig}
that the relative frequency of O-depleted stars in $\omega$ Cen is larger for [Fe/H] $\geq$ --1.2,
compared to the bulk of the population at [Fe/H] $\approx$ --1.75 dex.  Such an effect has already
been noted by, for example, \citet{ND95b} and \citet{CB10b}.

As regards [Na/Fe], again within the [Fe/H] interval common to both $\omega$~Cen and M22,
the range in [Na/Fe]  seen in both systems is comparable, of order 0.8--1 dex.  
This [Na/Fe] range is in agreement with that seen 
in most globular clusters \citep[e.g.,][]{CB10}.  Moreover, the lower [Na/Fe] 
values, which correspond to the O-rich stars, are similar between the two objects as well
as being similar to the [Na/Fe] values for halo field stars of comparable [Fe/H] \citep[e.g.,][]{JP09, CB10}.  
However, as noted by \citet{JP09} and \citet{CB10b}, at [Fe/H] $\geq$ --1.2,  stars with
notably larger enhancements in [Na/Fe] dominate in $\omega$ Cen, and there are relatively 
few stars with low values of [Na/Fe].   
In other words, using the terminology of \citet{CB09a}, the metal-rich population
of $\omega$~Cen is dominated by ``extreme'' stars while ``primordial'' stars are rare at these 
metallicities \citep[see also][]
{JP09, JP10}.  In contrast, M22, at least as regards the sample of \citet{AM09}, lacks such ``extreme'' stars. 
 
For [Al/Fe], if we again restrict ourselves to the [Fe/H] interval in which $\omega$ Cen and M22 have
stars in common, then we see results that are broadly similar to those for [Na/Fe].  In particular, a range
in [Al/Fe] of $\sim$1 dex is present in both systems though there is some indication that the M22
stars have a slightly lower ``primordial'' [Al/Fe] and a lower ``enhanced'' [Al/Fe] than do the 
$\omega$~Cen
stars.  This may be a result of different choices of analysis parameters rather than any real offset.  
The combined \citet{ND95b} and \citet{JP09} results for $\omega$~Cen in the bottom left panel of
Figure \ref{onaal_fig} present an intriguing picture.  For [Fe/H] values below approximately --1.2 dex,
it appears that the maximum [Al/Fe] value is constant at $\sim$1.0 dex, but at higher abundances,  
it decreases with increasing [Fe/H].  This effect contrasts with 
the situation for [Na/Fe] in $\omega$~Cen, where the highest values of [Na/Fe] occur at and above
[Fe/H] $\approx$ --1.2 dex.  \citet{JP09} discuss the possible implications of this result, noting it is certainly 
consistent with current AGB nucleosynthesis models, which predict that more Al is produced at
low metallicity, and more Na at higher metallicity, due to lower temperatures at the bottom of
the convective envelope, and shallower mixing, in more metal-rich stars \citep[e.g.,][]{VD08}.

\subsection{$r$- and $s$-process elements}

\begin{figure}
\begin{center}
\includegraphics[scale=0.42, angle=-90.]{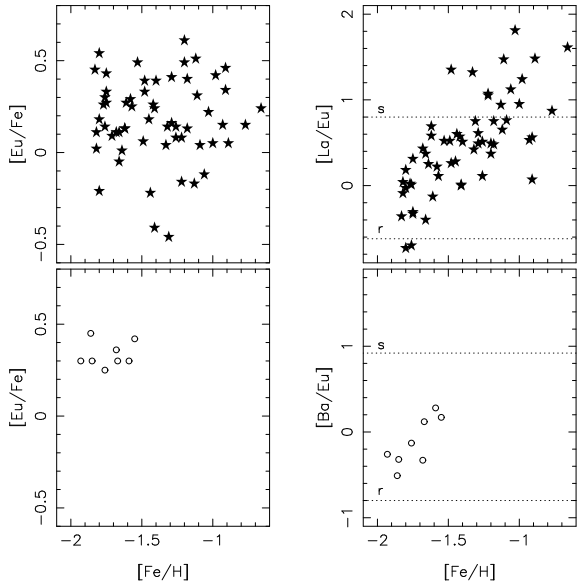}
\caption{{\it Left panels.} Europium-to-iron ratios as a function of [Fe/H]
for red giants in $\omega$ Cen (upper panel) and in M22
(lower panel). Note the absence of any clear trend of [Eu/Fe]
with [Fe/H] in both systems. {\it Right panels.} Lanthanum-to-Europium ratios as a function
of [Fe/H] for $\omega$ Cen red giants (upper panel) and Barium-to-Europium ratios as
a function of [Fe/H] for M22 stars (lower panel).  The dotted lines in the right panels
indicate the solar system  $r$-process ratio and an extreme $s$-process ratio taken
from \citet{McW97}.  
The increase of the ratios with [Fe/H] indicates the dominance of $s$-process contributions to 
the nucleosynthesis. The $\omega$ Cen data are from \citet{JP09}.  }
\label{eu_fig}
\end{center}
\end{figure}

We first examine the possible contribution of $r$-process nucleosynthesis in $\omega$~Cen and M22
by investigating the element-to-iron abundance ratio for Europium, an element whose abundance
is dominated by $r$-process nucleosynthesis.  The observational results are shown in the left panels 
of Figure
\ref{eu_fig} where the $\omega$ Cen data have been taken from \citet{JP09} as the \citet{ND95b}
data for Eu contains only upper limits for a limited sub-sample of stars.   Again the recent results
of \citet{JP10} are similar to those shown.  The vast majority of Galactic
globular clusters have [Eu/Fe] = +0.40 dex with relatively little cluster-to-cluster scatter \citep[e.g.,][]{RG04},
a value consistent with halo field stars of comparable [Fe/H] \citep{RG04}.
The value for the M22 stars in the lower left panel is clearly consistent with this value and there
is no indication of any intrinsic  variation in the [Eu/Fe] values.  The $\omega$ Cen values shown in the
upper left panel exhibit an apparently substantial scatter, though \citet{JP09} do not comment on it
indicating, presumably, that the scatter is primarily due to observational errors.  Certainly the small
number of $\omega$ Cen stars with apparently low [Eu/Fe] ratios are not discrepant in the [La/Eu]
plot shown in the upper right panel of Figure \ref{eu_fig}.  Specifically, only two of the seven stars
with [Eu/Fe] $\leq$ --0.1 in the upper left panel of Figure \ref{eu_fig} also have [La/Eu] $\geq$ 1.0
in the upper right panel of the Figure, while the other eight stars with [La/Eu] $\geq$ 1.0 all have
[Eu/Fe] values consistent with the bulk of the $\omega$ Cen sample.

\citet{JP09} and \citet{JP10} note  that the mean [Eu/Fe]
for the $\omega$ Cen stars is 0.1 to 0.2 dex lower than the value for other globular clusters, including M22,
and for the halo field.  Nevertheless, the lack of any variation of [Eu/Fe] with [Fe/H] in either $\omega$ Cen 
or M22 indicates that the $r$-process nucleosynthesis in these systems must be tightly coupled to 
that of Iron.  This is in contrast to the situation in the metal-poor globular cluster M15 where a distinct
range in [Eu/Fe] abundances, of order 0.5 dex, is seen at constant [Fe/H] \citep{SK97, Ot06}.

As regards the $s$-process contribution, we show in the upper right panel of Figure \ref{eu_fig} 
the abundance ratio [La/Eu] as a function of [Fe/H], again
using the data of \citet{JP09}.  Lanthanum is primarily synthesized by the $s$-process while, as noted 
above, Eu traces the $r$-process.  Their abundance ratio is therefore a measure of the relative 
importance of these two neutron capture nucleosynthetic processes \citep[see, e.g.,][]{McW97}.
The data reinforce the conclusions of \citet{ND95b}, \citet{VS00}, \citet{JP09} and
\citet{JP10} that as [Fe/H] increases in $\omega$ Cen, the $s$-process dominates the enrichment
of the neutron capture elements.

This appears also to be the case for M22.  The lower right panel of Figure \ref{eu_fig} shows a similar
abundance ratio versus [Fe/H] plot for the M22 red giants in the sample of \citet{AM09}.  
Here we have used 
Barium as the $s$-process tracer since \citet{AM09} did not measure La abundances.  While the 
number of M22 stars in the sample with measured abundances for both elements is relatively small,
the data nevertheless show a similar trend of increasing abundance ratio with increasing [Fe/H].
Indeed if we restrict the $\omega$ Cen data to only those stars with [Fe/H] $\leq$ --1.5, i.e., those
which overlap in [Fe/H]  with the M22 sample, then the slope of the ([La/Eu], [Fe/H]) relation is
essentially identical to the slope of the ([Ba/Eu], [Fe/H]) relation for  the M22 stars.  In other words
the rate of increase with [Fe/H] of the relative $s$-process contribution appears to have been the same
in both stellar systems, again suggesting that similar nucleosynthetic processes are at work. 

The high values of [Ba, La/Eu] at larger [Fe/H] in both M22 and $\omega$~Cen contrast 
with the situation in the majority of other globular clusters, where the ratio typically reveals a dominant 
$r$-process contribution.  \citet{RG04} list a mean [Ba, La/Eu] value of --0.23 $\pm$ 0.04 ($\sigma$ =
0.21) dex for 28 clusters with [Fe/H] values between --2.4 and --0.7 dex.  The one clear exception is
M4, for which \cite{II99} give $\langle$[Ba/Eu]$\rangle$ = +0.25 dex, indicative of a more substantive
$s$-process contribution to the gas from which the M4 stars formed.  We note, however, that there is no 
evidence for any intrinsic spread in the Barium (or Lanthanum) abundances in M4 \citep{II99,AM08}.

\begin{figure}
\begin{center}
\includegraphics[scale=0.43,angle=0.]{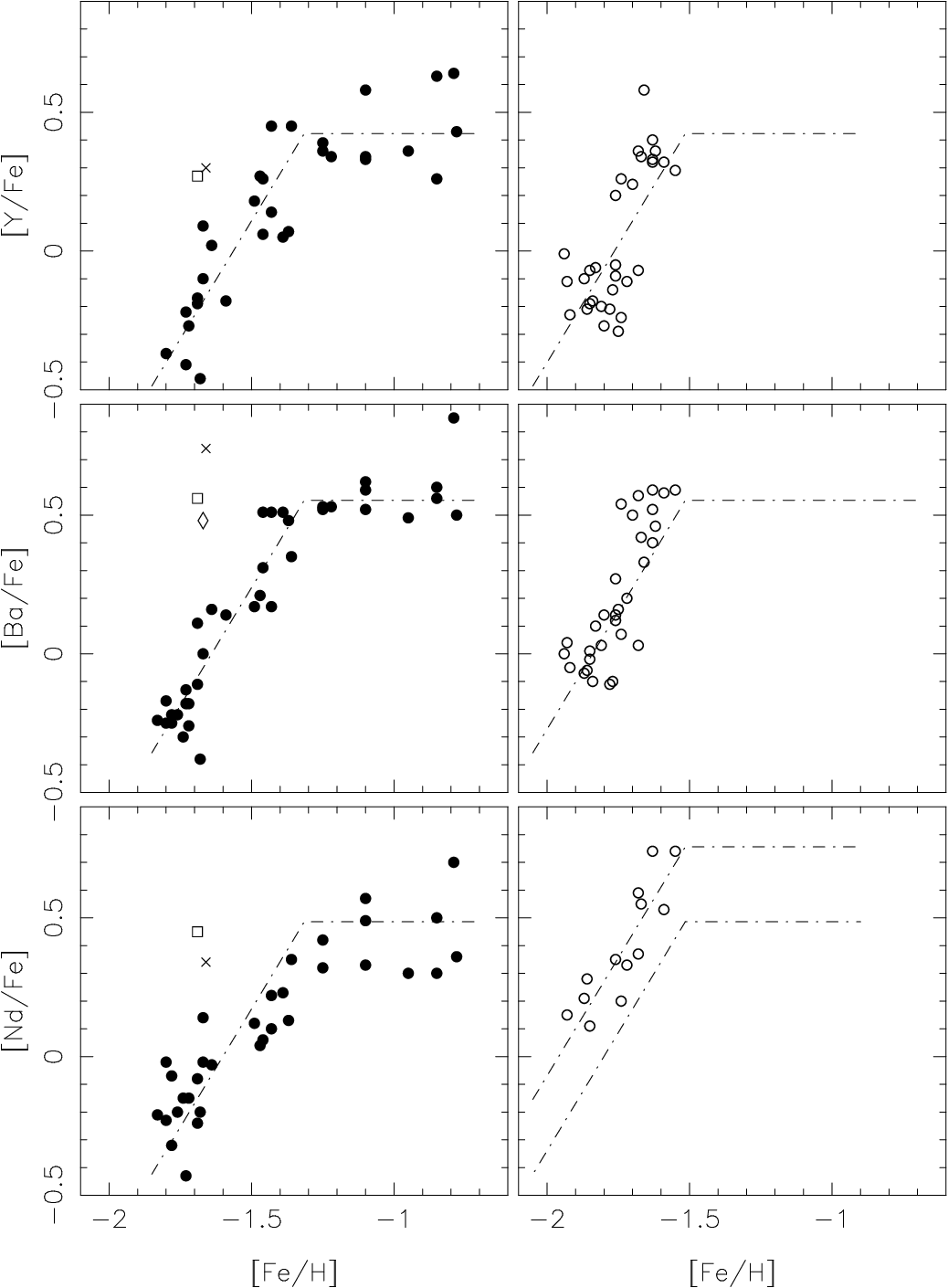}
\caption{
Element-to-Iron abundance ratios for the $s$-process elements Yttrium
(upper), Barium (middle) and Neodymium (lower panels) as a function of
[Fe/H] for $\omega$ Cen red giants (left panels) and for M22 red giants (right panels).  The 
dot-dash lines represent the relations adopted for the $\omega$ Cen stars; the slope of the linear
portion is the same for all 3 element ratios.  The open square is the CH-star ROA 279 while the
x-sign is the CN-strong star ROA 144 \citep{ND95b}.  These stars have been excluded from the fits, 
as was the [Ba/Fe] data for ROA 53, shown as the open diamond in the middle left panel. 
In the right panels the $\omega$ Cen relations have been translated to the M22 stars 
as described in the text. }
\label{yndba_fig}
\end{center}
\end{figure}

We now look at the $s$-process elements in more detail.  In the left panels of Figure \ref{yndba_fig} we
show the element-to-iron ratios for $\omega$ Cen red giants for the $s$-process elements Yttrium, 
Barium and Neodymium as a function of [Fe/H] using data from \citet{ND95b}.  As is evident from the 
figure, all three elements show similar behaviour.  There is an initial rapid rise in the element-to-iron abundance ratio, from a presumably $r$-process driven initial value, up to abundance ratios that are significantly above solar.  At higher metallicities, however, the increase ceases and the abundance
ratio remains constant as the Iron abundance continues to rise.  The total change in the abundance
ratios is of order 0.9 dex in all three cases.  

We have endeavoured to quantity 
this behaviour as follows.  First, we adopted [Fe/H] = --1.30 as the abundance above which the 
[Y, Ba, Nd/Fe] abundance ratios are constant.  We then used least squares to determine the slope
of the relation between the abundance ratio and [Fe/H] for the stars less abundant than the adopted
limit.  In carrying out the least squares fits, two stars, namely the extremely CN-strong star ROA 144
and the CH-star ROA 279 were excluded from the fits.  Both these stars have anomalously large 
[$s$-process/Fe] ratios for their [Fe/H] values \citep{ND95b} and may possess additional $s$-process
enhancements resulting from binary mass transfer.  ROA 53 was also excluded from the
([Ba/Fe], [Fe/H]) fit because of the apparently high abundance ratio for its [Fe/H], but this star is not
unusual as regards [Y/Fe] or [Nd/Fe], nor in [La/Fe] for that matter.  

The slopes found for each element 
ratio were all consistent with each other to within the uncertainties.  Consequently, the values were
averaged and a single slope of 1.70 dex/dex for $\Delta$[$s$/Fe]/$\Delta$[Fe/H] refitted to each of the 
element ratio datasets.  The value of the element ratio for the fitted relation at [Fe/H] = --1.3 was then 
extended to higher metallicities, and it provides a satisfactory representation of the data.  We note also
that the ([La/Fe], [Fe/H]) data of \citet{JP09} are consistent with these results.  Excluding the seven stars
with [La/Fe] values exceeding +1.3 dex, which may have had additional $s$-process enhancement
from mass transfer in binary star systems\footnote{We note that while mass transfer in binaries 
involving a thermally pulsing AGB star is often invoked to explain large $s$-process 
element enhancements, e.g., for halo field Ba~II stars and specifically by \citet{JP09} for the stars 
in $\omega$ Cen with high [La/Fe], the work of \citet{MM96, MM97} shows that in $\omega$ Cen, the
stars with large \mbox{$s$-process} enhancements appear generally to be single, unlike what is found
in the halo field.  In particular, while the multiple radial velocity observations of \citet{MM96, MM97} have
demonstrated that the $\omega$ Cen CH-stars ROA~55 and ROA~77 are indeed long period binaries, the
same data set shows that the stars ROA~421 and ROA~451 (stars 51132 and 39048 in \citet{JP09}),
which have very large [La/Fe] values \citep{JP09} and which are classified as Ba~II stars by 
\citet{TLE86}, are apparently single stars.  Hence adopting a ``mass transfer in binary system'' 
explanation for
the $\omega$ Cen red giants with large $s$-process element enhancements may require that
the binary is subsequently disrupted in the cluster environment. \citet{JP10}, however, note that
a proper accounting for hyperfine structure has revised downwards the [La/Fe] values of 
\citet{JP09}, especially for the most La-rich stars.  This reduces the need for invoking mass transfer
in binary systems.}, the distribution of the remaining 
stars is consistent with a 
linear increase in [La/Fe] with [Fe/H] for [Fe/H] $\leq$ --1.3, and the same slope of 1.7 dex/dex as for 
the \citet{ND95b} [Y, Ba, Nd/Fe] data, together with a constant value of [La/Fe] above that [Fe/H] value.
The data presented in \citet{JP10} are also, at least qualitatively, consistent with this interpretation.

In the right panels of Figure \ref{yndba_fig} we show the same $s$-process element to iron abundance
ratios as a function of [Fe/H] using the M22 data of \citet{AM09}.  In that dataset abundance ratios are
given from both the VLT/UVES and VLT/GIRAFFE samples for [Y/Fe] and [Ba/Fe] (21 stars) while the 
data for [Nd/Fe] come from the VLT/GIRAFFE observations only \citep[13 stars,][]{AM09}.  The M22 data 
show a distinct degree of similarity with those for $\omega$ Cen: again there is a steep rise in the
[$s$-process/Fe] element ratios with increasing [Fe/H].  The total range in [$s$-process/Fe] is 
$\sim$0.7 dex, slightly smaller than that seen in $\omega$ Cen.  M22 also does not show the `flat' part 
of the relation as it lacks stars of higher [Fe/H] when compared to $\omega$~Cen.  

To further investigate this similarity, we have endeavoured to transpose the ([$s$-process/Fe], [Fe/H])
relations found for $\omega$ Cen to M22.  The results of this are shown as the dot-dash lines in the
right panels of Figure \ref{yndba_fig}.  For [Y/Fe] and [Ba/Fe], the dot-dash lines shown for M22 are
exactly those for $\omega$ Cen save only that they have been shifted to lower [Fe/H] values by 0.2
dex.  In other words, a given [Y, Ba/Fe] value occurs at a [Fe/H] value in M22 that is 0.2 dex lower
than in $\omega$ Cen.  For [Nd/Fe], there are two dot-dash lines shown in the M22 panel.  The first, 
which has lower [Nd/Fe] at fixed [Fe/H], is the same case as for [Y/Fe] and [Ba/Fe], i.e., the $\omega$
Cen relation shifted to lower [Fe/H] values by 0.2 dex.  Clearly it is a poor representation of the 
M22 [Nd/Fe] data.  The second is identical to the first except that it has been shifted vertically by 0.27 dex in
[Nd/Fe] to match the M22 observations.  This systematic relative offset in the [Nd/Fe] versus [Fe/H]
relation between M22 and $\omega$ Cen may well have its origin in the Nd lines measured, and
$gf$-values adopted for those lines, rather than any true difference in [Nd/Fe] at equivalent [Fe/H]
values.  In particular, we note that  the [Nd/Fe] values for M22 come from a single Nd II line at 532nm 
\citep{AM09}
while those for $\omega$ Cen come from 3-6 Nd II lines, including the 532nm line \citep{NDT96}.

Intriguingly, the offset in [Fe/H] required to match the $\omega$ Cen and M22 [$s$/Fe]  relations is 
very similar to the difference between the peaks of the metallicity distribution functions for the two clusters, 
which is of order 0.1 dex (see Figure \ref{ghisto_fig}).  Indeed, when we
consider the uncertainties in the zero points of the abundance scales involved, the $\sim$0.2 dex offset
used in Figure \ref{yndba_fig}
is fully consistent with the $\sim$0.1 dex difference between the abundance distribution
peaks.  For example, we note that 
the M22 data in Figure \ref{ghisto_fig} come from \citet{DH09} who measured Ca triplet line strengths
calibrated to the [Fe/H] scale of \citet{KI03}.  The $\omega$ Cen data, on the other hand, come from
\citet{NFM96} who measured Ca triplet and Ca K line strengths and calibrated principally to
the [Ca/H] values of \citet{ND95b}.  A mean [Ca/Fe] value was then applied to generate the [Fe/H]
distribution.  Given these different approaches, it is indeed likely that the difference in abundance 
distribution peaks seen in Figure \ref{ghisto_fig}
is uncertain at the $\sim$0.1 dex level.  We note further that even with the most recent work, the 
absolute scale of globular cluster abundances
remains uncertain at the $\sim$0.1 dex level.  For example, see the discussion of the various abundance
scales in \citet{CB09}.

Nevertheless, the similarity in the slopes of the ([$s$/Fe], [Fe/H]) relations, and
the similar overall range in [$s$/Fe] values, argues rather strongly that the $s$-process 
nucleosynthesis process involved was very similar in both M22 and $\omega$ Cen.  Further,
it is worth noting that the slope for Yttrium, which is a first $s$-process peak (or $light$-$s$) element, 
is evidently the same as for the second $s$-process peak (or $heavy$-$s$) elements Ba, Nd (and La). 
Moreover, the abundance ratio [Ba/Y], which is an [$hs$/$ls$] indicator, shows  no dependence
on [Fe/H] in either $\omega$ Cen or M22, and takes the same value, 0.18 $\pm$ 0.04 (std error of
the mean) in both clusters, again emphasising the similarity of the $s$-process nucleosynthesis in
these systems.
This lack of any dependence on [Fe/H] of the [Ba/Y] ratio, despite the substantial changes in both
[Y/Fe] and [Ba/Fe] with the [Fe/H], shows that the mechanism contributing the $s$-process elements
changed only as the regards the amount of $s$-process elements produced, and not in any other
significant way, as the Iron abundance increased.

The mechanism for generating $s$-process elements is
most likely relatively low-mass thermally pulsing AGB stars \citep[e.g.,][]{BGW99}.
It appears then that there was an epoch in both $\omega$ Cen and M22 when the 
production of $s$-process elements exceeded the on-going production of Iron so that the [$s$/Fe]
ratios increased substantially from low (pre-cluster $r$-process?) values to significantly in 
excess of the solar ratios.  In M22 it would appear that the star formation then ceased while in
$\omega$ Cen it continued, but without the excess $s$-process element  production, as the [$s$/Fe] 
values now remain constant as [Fe/H] rises.

\section{Discussion}

The nucleosynthetic processes that are likely to have occurred in $\omega$ Cen are outlined in detail
in, for example, \citet{JP09}, \citet{JP10} and \citet{Ro07, Ro10}, and need not be discussed further here.  
Rather the 
emphasis in this contribution is on the significant degree of similarity between $\omega$ Cen and 
M22 in their element-to-Iron ratios as a function of [Fe/H].  This similarity, particularly for the $s$-process 
elements, suggests strongly that similar chemical enrichment processes occurred in both systems.  
As a result, any explanation for the chemical abundances, such as an origin in the nucleus of a 
now disrupted dwarf galaxy, must apply to the other if postulated for one.  Consequently, the difference 
in mass between M22 and $\omega$ Cen (and for that matter also with M54, which has a comparable 
luminosity to $\omega$ Cen) is intriguing. Total mass is an important parameter for retaining 
gas from which to form successive generations of stars and so it is worthwhile to ask if it is possible that 
M22 could have been significantly more massive in the past.  To investigate this question
we assume that the M22 dwarf 
galaxy progenitor was disrupted at early times, as is the case for the $\omega$~Cen dwarf galaxy
progenitor in the models of \citet{BF03}.  Thus we can compare the subsequent dynamical evolution
of M22 with that for halo globular clusters. Under this assumption, it does not seem likely that M22
could have lost a substantial amount of stellar mass.

For example, in the compilation of \citet{DG99}, the orbit of M22 is typical for inner halo objects 
\citep[e.g.,][]{CB07}.  It is prograde with $\Theta$ =  178 $\pm$ 20 km s$^{-1}$,  apo-
and pericentric distances of approximately 9.5 and 2.9 kpc, respectively, and a period of 
$\sim$200 Myr \citep{DG99}.  This suggests that M22 is unlikely to be strongly affected dynamically
by disk and bulge shocks.
\citet{GO97} reach similar conclusions, listing for M22 a destruction rate from disk and bulge
shocks of $\sim$0.3 inverse Hubble times, i.e., a destruction timescale exceeding 3 Hubble times.  
Further, with a half-mass 2-body relaxation time 
of order 1.4 Gyr \citep{WH96, GO97}, M22 is also unlikely to have lost significant mass through the 
evaporation of stars.
Indeed the parameters for M22 place it inside the `survival triangles' or `vital diagrams' 
shown in \citet{GO97}.  Thus
it seems improbable that M22 could have lost the factor of 5 or more by which its present day mass differs
from that of $\omega$ Cen.  

In this context it is then interesting to speculate that perhaps the M22 dwarf galaxy progenitor was
of lower mass total mass than that postulated for the $\omega$~Cen system and that of the Sgr dwarf
which currently has M54 as its central star cluster.  This lower total mass for the original M22 system 
might then have meant that it did not retain gas in the central regions as efficiently, given the shallower
potential well.  The possible consequences could then be less mass built-up in the central star
cluster and insufficient time to allow the star formation and chemical evolution to evolve to the higher 
[Fe/H]  values seen in $\omega$ Cen.  Similarly, in the scenario advanced for the ultimate fate of M54 
\citep[e.g.,][]{CB10b}, the M22 progenitor would not have had sufficient mass to acquire metal-rich stars 
from the nucleus of the dwarf galaxy in the tidal disruption process.

We note for completeness that Terzan 5, the other globular cluster with a significant internal [Fe/H]
abundance range \citep{FD09}, has recently been shown \citep{LF10} to be more massive than 
previously thought: \citet{LF10} estimate the mass as 2 $\times$ 10$^{6}$ solar masses.   This is 
comparable to that for $\omega$ Cen for which mass estimates range up to 
5 $\times$ 10$^{6}$ M$_{sun}$ \citep[][see also \citet{VV06}]{MM95}. Further, as a metal-rich object in the Galactic Bulge,
Terzan 5 is likely to be strongly affected by tidal shocks and may therefore have been yet more massive
in the past.

\begin{figure}
\begin{center}
\includegraphics[scale=0.43, angle=0.]{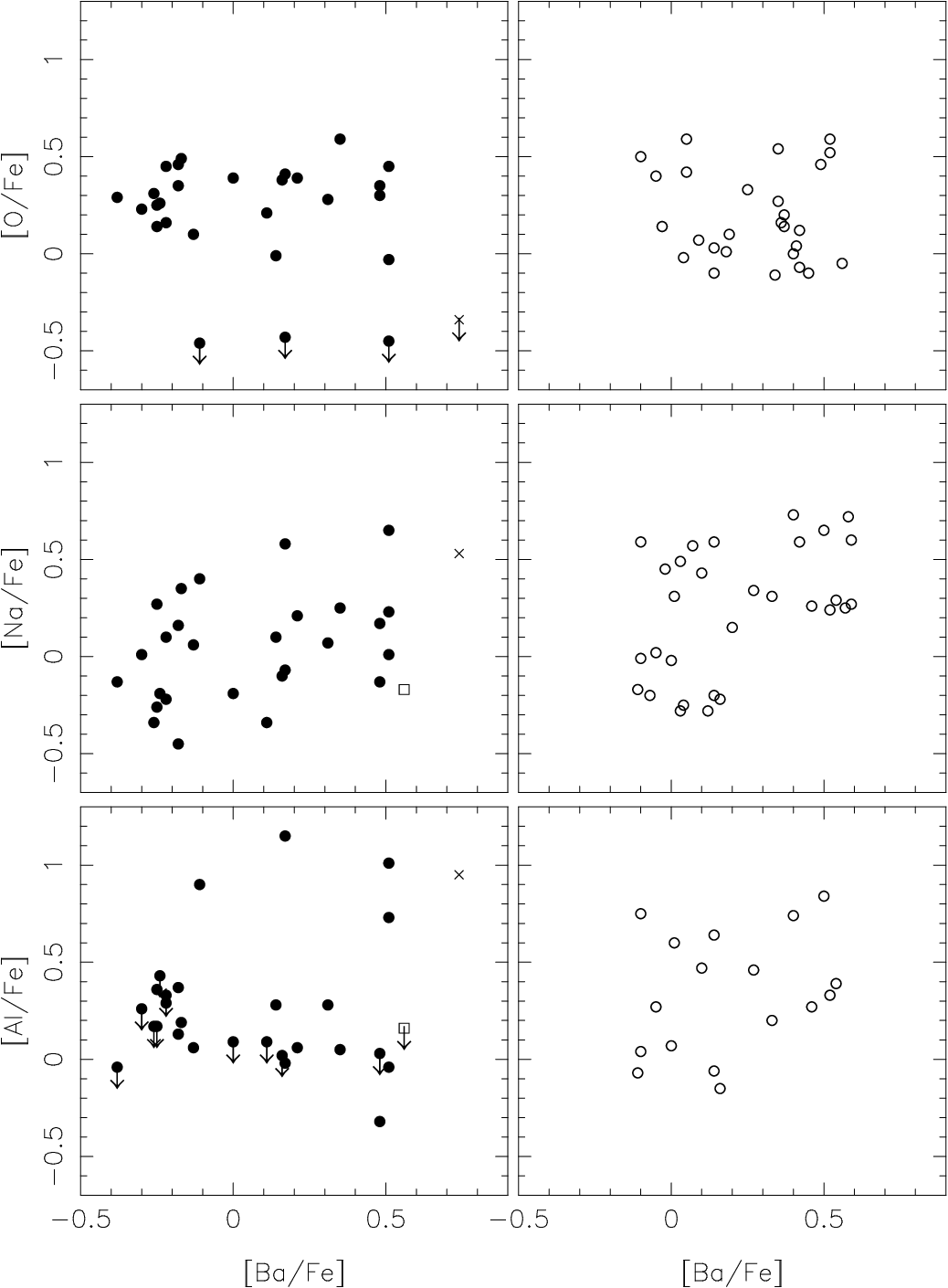}
\caption{Oxygen-to-Iron (upper panels), Sodium-to-Iron (middle panels) and Aluminium-to-Iron 
(lower panels) abundance ratios as a function of [Ba/Fe]  for red giants in
$\omega$ Cen (left panels) and M22 (right panels).  The $\omega$ Cen red giants shown all have 
[Fe/H] $\leq$ --1.3 dex, the [Fe/H] range over which the [Ba/Fe] ratio changes significantly
(see Figure \ref{yndba_fig}).  As in Figure \ref{yndba_fig}, the open square in the $\omega$ Cen
panels is the CH-star ROA 279 while the x-sign is the CN-strong star ROA 144.  The lack of any 
correlation between these abundance ratios suggests that in both stellar systems the nucleosynthesis 
process that generates the $s$-process element enrichment is distinct from that which involves the 
O-Na anti-correlation.}
\label{ba_al_fig}
\end{center}
\end{figure}

One final question that the data presented here allow us to ask is whether there is any relation 
between the process(es) that govern the O-Na anti-correlation and those that govern the $s$-process 
element enhancements.  The former is seen in $\omega$ Cen and M22 \citep{ND95a,AM09}, and 
essentially all globular clusters \citep[e.g.,][]{CB10}, while the latter is seen in $\omega$ Cen and M22 
as well as in a very
small number of other clusters.  For example, \citet{DY05} and \citet{DY08} have shown that there are 
small intrinsic variations in $s$-process element to Iron abundance ratios in the clusters NGC~6752
($\Delta$[Y, Zr, Ba/Fe] $\approx$ 0.1 dex) and NGC~1851 ($\Delta$[Zr, La/Fe] $\approx$ 0.3 dex).  These 
variations appear to correlate with [Na/Fe] and  [Al/Fe].  No [Fe/H] variations, however, were detected.
Nevertheless, \citet{CG10} have recently presented evidence for the existence of two different groups
of stars in NGC~1851 whose Iron abundances differ by 0.06--0.08 dex.  Both components show the
O-Na anti-correlation and there is also a hint that the more metal-rich stars have slightly higher
[$s$/Fe] values \citep{CG10}.  A larger sample of NGC~1851 stars is required to establish if this is a 
real effect and if the slope of any ([$s$/Fe], [Fe/H]) relation is similar to that seem in $\omega$ Cen 
and M22. 

A relation between the O-Na and $s$-process effects might be expected given that 
thermally pulsing AGB stars are the main source of $s$-process elements
\citep[e.g.,][]{BGW99} and, such stars, albeit of somewhat higher mass, are also often invoked 
to explain the O-Na anti-correlation \citep[e.g.,][]{DV07}.
The relatively large range in [$s$/Fe] seen in both $\omega$ Cen and M22 thus makes them ideal 
objects with which to investigate this question.

In Figure \ref{ba_al_fig} we show the [O/Fe], [Na/Fe] and [Al/Fe] abundance ratios as a function of
[Ba/Fe] for $\omega$ Cen (left panels) and M22 (right panels) red giants.  For $\omega$ Cen, 
we have plotted only those stars with [Fe/H] $\leq$ \mbox{--1.3}, since below this value the abundance 
ratio changes
significantly while above it the [Ba/Fe] ratio is essentially constant (see Figure \ref{yndba_fig}).  
Stars over the entire observed [Fe/H] range are shown for M22.  We note
for completeness, however, that while the majority of the $\omega$ Cen metal-rich stars are O-poor and
Na, Al-rich, as seen in Figure \ref{onaal_fig}, there do exist some metal-rich stars that are O-rich and 
Na, Al-poor.
Thus there is a range in [O/Fe], [Na/Fe] and [Al/Fe] at the essentially fixed (but high) [Ba/Fe] value
these stars possess.  

Both the right and left panels of Figure \ref{ba_al_fig} do not present any 
compelling case for a correlation between the [O/Fe], [Na/Fe], and [Al/Fe] abundance ratios with
[Ba/Fe], in line with the conclusions of \citet{GS08}.  At best there is a suggestion of a trend in 
which [Na/Fe] might be slightly higher at larger
[Ba/Fe] values, though the [O/Fe] and [Al/Fe] values do not show any comparable trend.  Plots of
[Na/Fe] and [Al/Fe] against [La/Fe] for $\omega$ Cen red giants using the data of \citet{JP09}
show the same result --- there is some indication of a trend for increasing [Na/Fe] with increasing
[La/Fe] but the [Al/Fe] data are consistent with no correlation with [La/Fe].   The results of \citet{JP10}
appear similar.

Thus is seems apparent
that the process(es) which generate the O-Na anti-correlation must be relatively distinct from those that
generate the $s$-process element enhancements.  This is despite the fact that in both $\omega$ Cen 
and M22 the existence of a range in [Na/Fe] at fixed [Ba/Fe], and vice versa, suggests that both
processes were occurring together.  It is hard to see how this can be the case invoking only AGB stars
of different mass in a closed system -- it is more 
likely that gas flows into and out of the star forming systems are required as suggested, for example,
by \cite{BN06} and \citet{Ro10}.  A detailed comparison of the abundance patterns in the Sgr dwarf 
galaxy central star cluster M54 with those of $\omega$ Cen and M22 would certainly assist in 
investigating this possibility.

\section*{Acknowledgments} 


The authors are grateful to Prof John Norris and Dr David Yong for their input to the original version
of the manuscript and to the referee for their comments.


\end{document}